\documentclass[10pt,twocolumn,letterpaper]{article}

\usepackage{ijcb}
\usepackage{times}
\usepackage{epsfig}
\usepackage{graphicx}
\usepackage{amsmath}
\usepackage{amssymb}

\usepackage[english]{babel}
\usepackage{subcaption}
\usepackage{mathtools}
\usepackage{amsthm}
\usepackage{amssymb}
\usepackage{multirow}
\usepackage{pifont}
\usepackage{makecell}
\usepackage{adjustbox}

\ijcbfinalcopy

\ifijcbfinal\fi
\begin{document}

\title{Formalizing PQRST Complex in Accelerometer-based Gait Cycle for Authentication}

\author{Frank (Sicong) Chen\\
Syracuse University \\
900 S Crouse Ave\\
Syracuse, New York, 13210\\
{\tt\small schen154@syr.edu}
\and
Amith K. Belman\\
California State University \\
5241 N. Maple Ave. \\
Fresno, California, 93740 \\
{\tt\small akamathbelman@mailcsufresno.edu}
\and
Vir V. Phoha\\
Syracuse University \\
900 S Crouse Ave\\
Syracuse, New York, 13210\\
{\tt\small vvphoha@syr.edu}
}

\maketitle
\thispagestyle{empty}

\begin{abstract}
    Accelerometer signals generated through gait present a new frontier of human interface with mobile devices. Gait cycle detection based on these signals has applications in various areas, including authentication, health monitoring, and activity detection. 
    Template-based studies focus on how the entire gait cycle represents walking patterns, but these are compute-intensive. Aggregate feature-based studies extract features in the time domain and frequency domain from the entire gait cycle to reduce the number of features. However, these methods may miss critical structural information needed to appropriately represent the intricacies of walking patterns. 
    To the best of our knowledge, no study has formally proposed a structure to capture variations within gait cycles or phases from accelerometer readings. We propose a new structure named the PQRST Complex, which corresponds to the swing phase in a gait cycle and matches the foot movements during this phase, thus capturing the changes in foot position. In our experiments, based on the nine features derived from this structure, the accelerometer-based gait authentication system outperforms many state-of-the-art gait cycle-based authentication systems. Our work opens up a new paradigm of capturing the structure of gait and opens multiple areas of research and practice using gait analogous to the "QRS complex" structure of ECG signals related to the heart. 
\end{abstract}

\section{Introduction}

    Gait analysis, focusing on the pattern of body movement during walking or running, has received increasing attention in recent years \cite{gafurov2007survey, wan2018survey}. There are two popular approaches used to analyze gait data. The first is video-based, which uses external cameras to record videos that directly capture body movements \cite{lee2002gait, niyogi1994analyzing}. The second is sensor-based, which analyzes signals from sensors that are attached to the human's body \cite{hou2014systemforhumangait}. Accelerometers, as a common built-in sensor in mobile devices, have become one of the most popular sensors in sensor-based gait analysis \cite{tamviruzzaman2009epetwhen}.
    
    Prior studies \cite{gafurov2007gaitauthen, mantyjarvi2005identifyingusers} have found that there are repetitive patterns in accelerometer signals that correspond  to two foot steps and constitute a gait cycle. These studies presented methods to detect the gait cycle, created templates or extracted features from the cycle, and matched feature vectors or templates for recognition. However, using the entire gait cycle to create templates requires much training data and a lot of computation time, which is less efficient for mobile devices. Also, aggregate gait features, such as the minimum and maximum of the whole gait cycle, cycle length, and velocity \cite{anwary2018automatic}, miss critical structural information of gait cycles. Thus, they may not represent gait information precisely and are easy to be imitated. Some studies \cite{bejarano2014novel, Maqbool2017realtime} associated "Heel contact" and "Toe off", which are two essential foot events that occur within the gait cycle, with accelerometer signals to help detect gait cycles and extract more accurate features. These two events divide the gait cycle into the stance phase, where accelerometer signals remain almost constant, and the swing phase, where accelerometer signals vary greatly. However, these studies didn't explain why these events match the points in accelerometer readings, what foot events occurred during the swing phase, and why accelerometer signals vary a lot in this period. We define a structure that consists of the swing phase, including the raising of the foot, flight of the foot, and lowering of the foot, and relate it to the changes in accelerometer signals. This structure appears in every gait cycle. 
    
    The main contributions of this paper are as follows:
    \begin{itemize}\label{contribution}
        \item We formally define a standard structure within gait cycles from accelerometer signals, called PQRST Complex. 
        It captures the most representative and persistent part of the walking patterns because the rate of change in accelerometer readings in the swing phase is very high as compared to the stance phase. 
        \item We present nine features using the PQRST Complex that capture the gait cycle's change points, thus capturing variations within the cycle. 
        \item We achieved an average correct classification rate (CCR)\label{abbr:CCR} of 90.51\% on an authentication system using PQRST Complex with performance better than most previous research with a smaller set of features.
    \end{itemize}
    
    The rest of this paper is organized as follows. Section \ref{sec:relatedwork} shows prior works in this field. In Section \ref{sec:PQRSTcomplex}, we describe the definition of the PQRST Complex and analyze features extracted from it in detail. The performance evaluation is displayed in Section \ref{sec:exp}. Finally, we conclude our work in Section \ref{sec:conclusion}.

\section{Related work}\label{sec:relatedwork}

    Signals from the accelerometer readings can be used to analyze gait \cite{ailisto2005identifying, annadhorai2008human, derawi2012smartphones}. Several methods for gait cycle detection and recognition have been proposed in the last decade. Gafurov \etal \cite{gafurov2007gaitauthen, gafurov2007spoof, gafurov2006robustness} proposed a straightforward method for gait cycle detection with an assumption that the minimas in accelerometer signals are the starting points and ending points for each gait cycle, and each cycle contains two steps. They used six consecutive cycles and used the median points of these cycles to create a feature vector and achieved a 86.3\% recognition rate. Derawi \etal \cite{Derawi2010Improved} presented an improved cycle detection method on accelerometer signals. Besides using the same assumption as Gafurov \etal \cite{gafurov2007gaitauthen}, they combined estimated cycle length (the length of one gait cycle usually varies between 800 ms - 1400 ms) and Neighbour Search to find all minimum points and utilized all detected cycles to create templates. They proposed the cyclic rotation metric (CRM)\label{abbr:CRM} to measure the distance between input samples and templates for authentication. The lowest Equal Error Rate (EER) they achieved was 5.7\%. Rong \etal \cite{Rong2007Identification} presented a gait cycle partition method and gait recognition algorithm based on Dynamic Time Warping (DTW)\label{abbr:DTW}. They used the first zero value point right after the local minimum point as the starting and ending points to separate gait cycles, and defined one gait cycle consisting of four consecutive points of zero value. They applied a time-warping network on the gait cycle they defined to create a gait feature code for each cycle, and used it for their gait recognition algorithm. They got 6.7\% EER in their recognition system. All of the methods mentioned above considered one gait cycle as a whole. They created templates from detected cycles and measured the distance between cycles from input samples and templates by using similarity-based or distance-based methods. Although they achieved relatively low EER in gait recognition or authentication, it was time-consuming to compare distance between template as each template consists of around 100 points. Moreover, none of them have given a standard structure of gait cycles. They only detected the gait cycle with starting points and end points without caring about the exact patterns that gait cycles contain. 
    On the other hand, aggregate feature-based studies used features extracted from gait cycles to create feature vectors. Alobaidi \etal \cite{alobaidi2022real} extracted more than 300 features in both time domain and frequency domain, including mean, standard deviation, median, etc., of the gait cycle for activity classification. They got an average EER of 11.38\% in recognizing normal walking activity.

    As Neumann \cite{neumann2010kinesiology} illustrated in his book, human walking consists of a cyclic series of movements. In general, the gait cycle begins with the foot (heel) contacting the ground, continues as the foot (toe) leaving the ground, and ends with the same foot (heel) contacting the ground again. Thus, the beginning and end of the gait cycle are usually referred to as heel contact (HC). The toe leaving the ground separates the gait cycle into two phases: the stance phase and the swing phase. It makes these two foot events crucial for gait cycle detection. Some studies \cite{bejarano2014novel,Maqbool2017realtime} found these two events could be located in accelerometer signals to help them detect the gait cycle and estimate velocity. However, some other small foot events may also affect the changes of accelerometer signals, such as the raising of the foot, flight of the foot, and lowering of the foot during the swing phase. To the best of our knowledge, only a few studies have investigated these small events within the gait cycle. Anwary \etal \cite{anwary2021Insolebased, anwary2018automatic} found that eight foot events can be located in accelerometer signal collected from an accelerometer attached to shoes. They observed that the part where accelerometer signals vary greatly corresponds to the swing phase, and they used this observation to separate the stance and swing phases for step detection. Their results showed that this part of acceleration signal could be used to estimate the swing phase, step length, and velocity. However, they only focused on the terminal stance and toe-off events detection. They did not analyze the structure of these changes and how these foot events relate to the signal, so the gait events they separated were not accurate. In addition, accelerometer signals vary significantly in the swing phase and remain almost constant in the stance phase, and the swing phase only accounts for 40\% of the whole cycle. It suggests that it may be possible to discard 60\% of the gait cycle (stance phase) and focus on only 40\% of the gait cycle (swing phase) to improve the recognition efficiency. The structure we captured matches the swing phase. It captures the foot movements during this period, and we give an explicit explanation of why and how they are matched. 

    Many accelerometer-based gait authentication systems have been proposed in recent years and got promising performance. Mufandaidza \etal \cite{mufandaidza2018continuous} developed a smartphone user authentication system using raw accelerometer data. They used the DTW algorithm to synchronize the accelerometer data from all three axes and fed them into a feed-forward neural network. The best True Positive Rate (TPR)\label{abbr:TPR} they got was 73.68\%, and the lowest False Positive Rate (FPR)\label{abbr:FPR} they got was 22.22\%. Some studies have proposed gait cycle based authentication systems. Thang \etal \cite{thang2012gait} presented two approaches for identification based on gait using accelerometer data. Their first approach extracted average gait cycles, which they called time domain features, and used DTW for comparison. The average accuracy they got was 79.1\%. Their second approach used Fast Fourier Transform (FFT) to extract frequency domain features and used Support Vector Machine (SVM) to classify feature vectors. They got an average CCR of 92.7\%. Nickel \cite{nickel2012accelerometer} constructed an accelerometer-based gait authentication on the smartphone. They came up with a cycle-based method, which used raw gait cycle data combined with some features extracted by fixed-length time interval segmentation. They used some machine learning algorithms for classification, and their methods had a low EER, which was around 6\% to 7\%.

\section{Capturing the PQRST Complex structure from accelerometer signal in gait} \label{sec:PQRSTcomplex}
    
\subsection{Capturing the PQRST Complex structure in accelerometer signal}
    
    \begin{figure}[ht]
        \centering
        \includegraphics[width=\columnwidth]{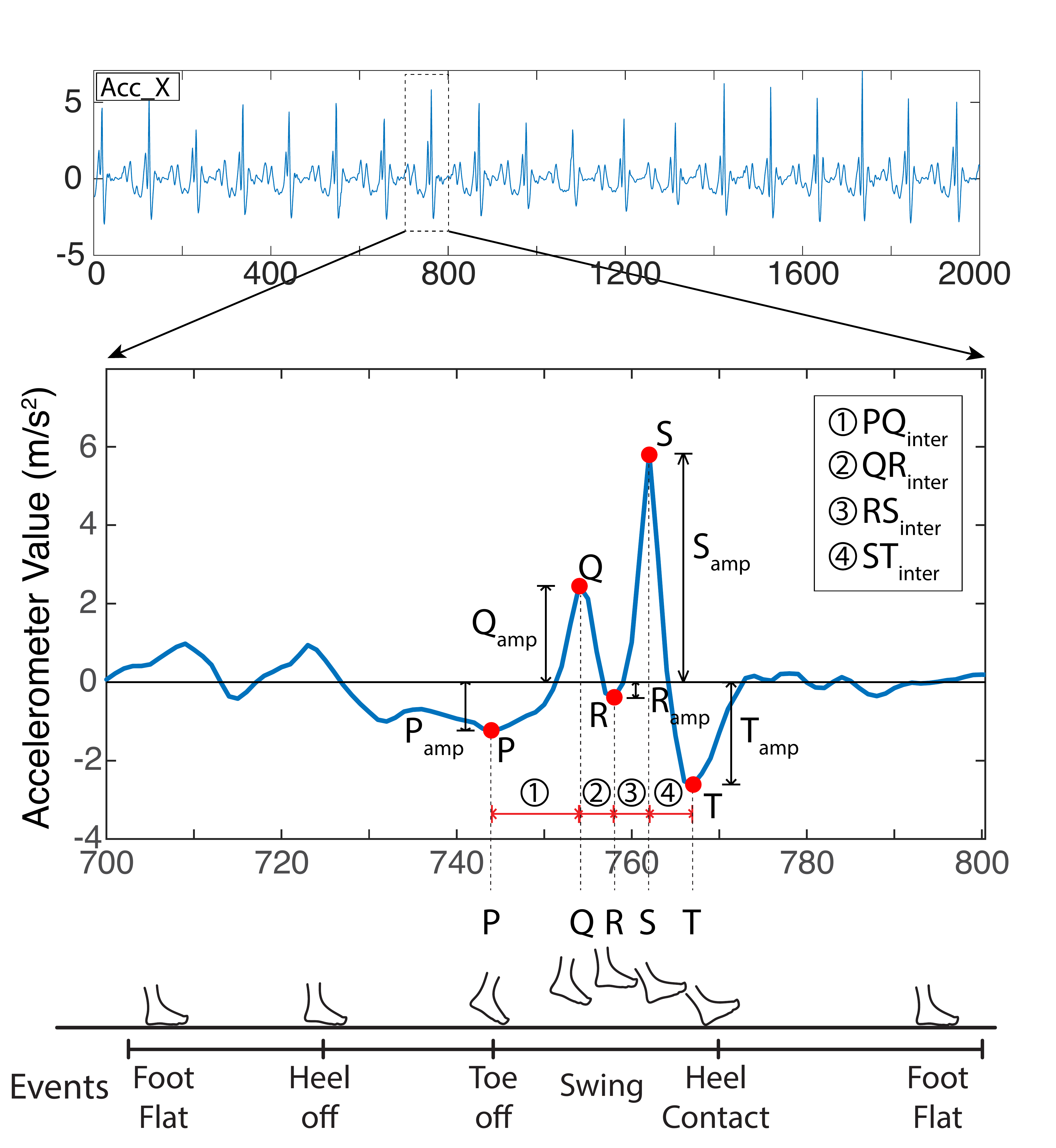}%
        \caption{\textbf{(Top):} Accelerometer signal ("Acc\_X") plot from subject 3.  \textbf{(Middle):} One of the repetitive patterns of accelerometer signal ("Acc\_X") plot from subject 3. The PQRST Complex is captured and marked by red points. Features that we extract from this structure are also marked. \textbf{(Bottom):} Events, which one gait walking cycle includes, match the accelerometer signal. The PQRST Complex corresponds to the swing phase during the walking cycle.}
        \label{fig:accX2PQRST}
    \end{figure}

    The accelerometer has three axes, in the rest of this paper, we use "Acc\_X" to represent the data of the x-axis of the accelerometer, and "Acc\_Y" and "Acc\_Z" are used to represent the data of the y-axis and the z-axis of the accelerometer respectively. The x-axis of accelerometer signal plots are the time with unit 10 ms, and the y-axis is the acceleration value with the unit $m/s^2$.
    
    \subsubsection{How is PQRST Complex defined?}
    
   We explain the following structure on a representative subject -- subject 3 from our dataset. Figure \ref{fig:accX2PQRST} (Top) displays the plot for "Acc\_X" from subject 3. It shows repetitive patterns in the accelerometer signal. As Mantyjarvi \etal \cite{mantyjarvi2005identifyingusers} presented, these repetitive patterns represent walking cycles. To take a closer look at the structure of each walking cycle, we zoom in on one of the repetitive cycles and put it in the middle part of Figure \ref{fig:accX2PQRST}. As Anwary \etal \cite{anwary2021Insolebased} mentioned in their work, signals from the accelerometer attached to shoes are relatively steady in the stance phase when compared to the swing phase. In our study, we use accelerometer signals from phones that are placed in pants pocket. We observe that these signals also have a steady part corresponding to the stance phase and a fluctuating part corresponding to the swing phase. Figure \ref{fig:accX2PQRST} (Middle) illustrates this observation. The middle portion shows the swing phase, on which we define the PQRST complex.
    Signals before the point $P$ and after the point $T$ do not fluctuate much. It means that they match the stance phase since the foot that contacts the ground is stationary to support the body in this period. Point $P$ corresponds to the "Toe off" event. After this point, the foot starts to swing and enters the swing phase. The swing phase will end when the event "Heel contact" takes place, and the point $T$ matches this event. Since acceleration values vary a lot between $P$ and $T$, intuitively, this section of the signal would be a rich source of information for feature extraction and authentication.  
    
    According to the variation of the acceleration values, the swing phase can be divided into several short periods. After the point $P$, the foot starts moving forward and upward, which leads to a speed up. It matches the $PQ$ part, where the acceleration increases. As the foot moves forward and upward, it will eventually become flat in the air and reach the highest position. During this process, the acceleration decreases and matches the $QR$ part. After that, the foot still moves forward but starts moving downward. It speeds up, leading to an increase in acceleration, corresponding to the $RS$ part. Finally, before the foot contacts the ground, it has to slow down, and the acceleration decreases, which matches the $ST$ part. After the point $T$, there is still a part of the signal where the acceleration increases. It starts from some negative values and increases to around zero. The reason for this is, after the "heel contact" event the entire foot will eventually fully contact the ground, and the acceleration will be around zero. After the "Heel contact" event, which corresponds to the point $T$, the gait cycle enters into the stance phase, and the accelerometer value does not have too many fluctuations until the stance phase ends. At the bottom of Figure \ref{fig:accX2PQRST}, we display the matching relationship between the accelerometer signal and foot events within the gait cycle. 
    
    \begin{figure}[!h]
    \begin{minipage}{0.49\linewidth}
    \centering
    \subfloat[Plot of accelerometer signal ("Acc\_Y") from subject 3.]{
        \label{fig:PQRSTY}
        \includegraphics[width=\linewidth]{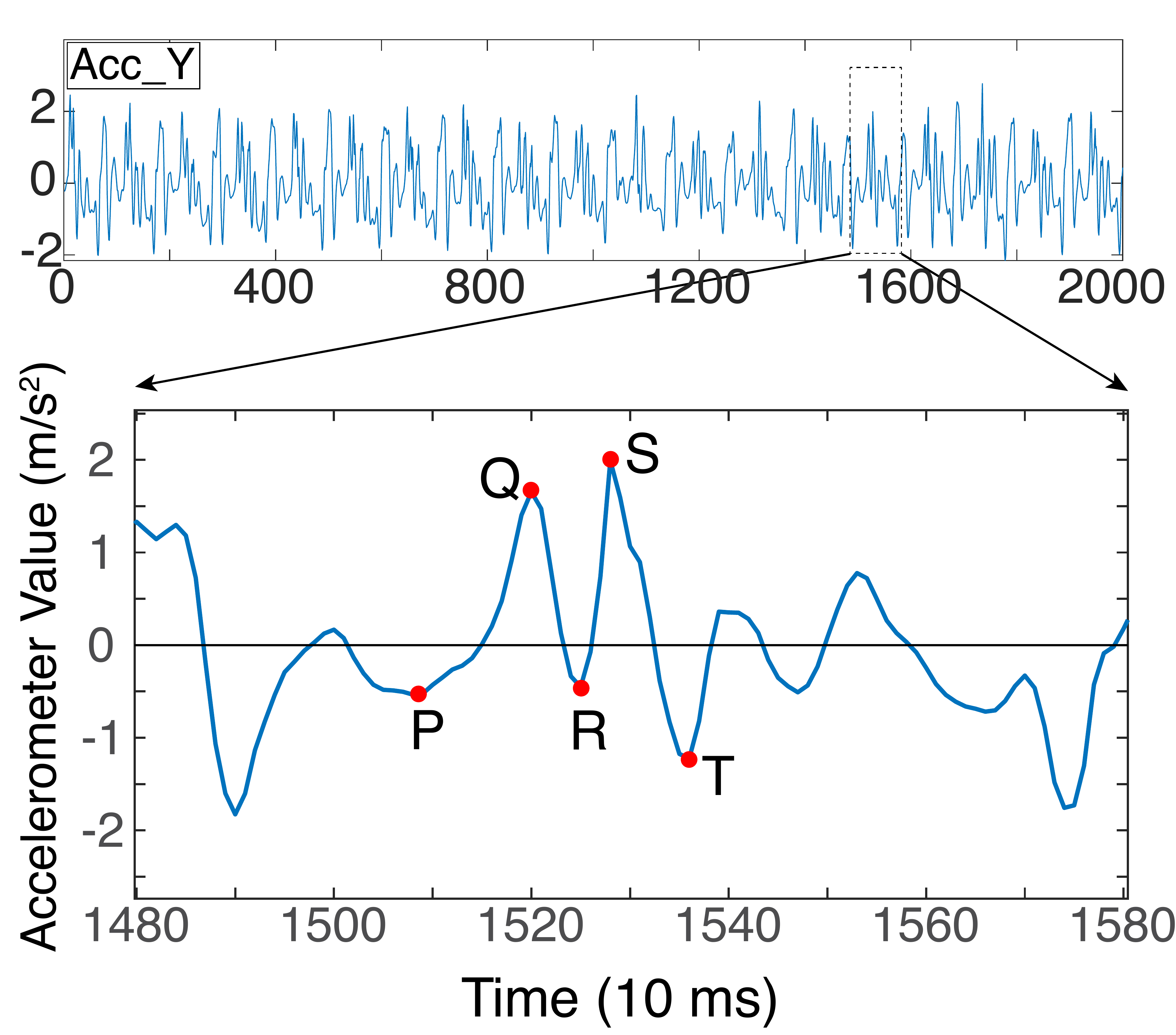}}
    \end{minipage}%
    \begin{minipage}{0.49\linewidth}
    \centering
    \subfloat[Plot of accelerometer signal ("Acc\_Z") from subject 3.]{%
        \label{fig:PQRSTZ}
          \includegraphics[width=\linewidth]{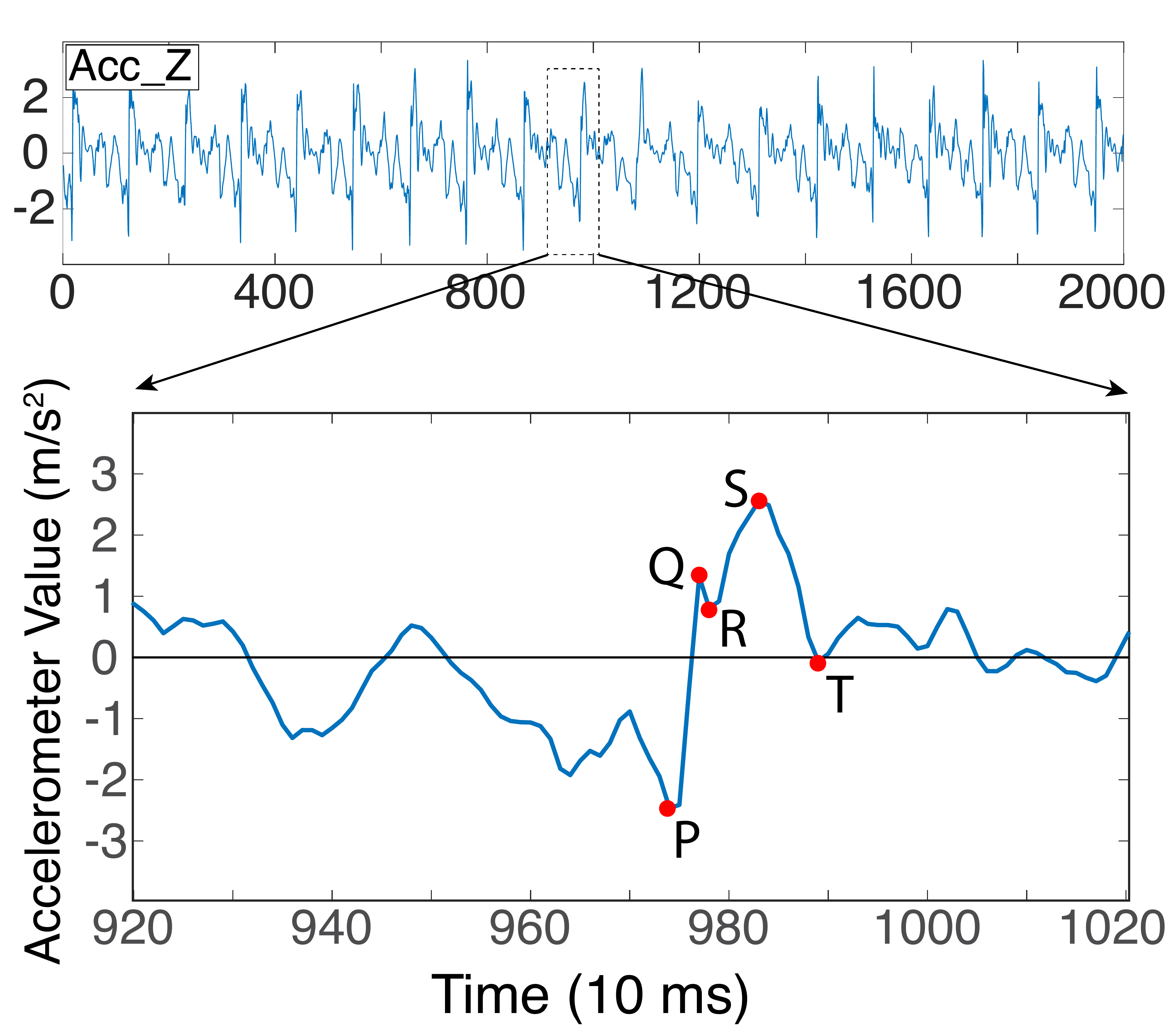}%
        }
    \end{minipage}%
    \caption{\textbf{(Left):} Accelerometer signal ("Acc\_Y") plot. \textbf{(Right):} Accelerometer signal ("Acc\_Z") plot from subject 3. For each figure, the accelerometer signal plot is on the top, and one of the repetitive patterns of accelerometer signal with annotation of PQRST Complex is on the bottom. }
    \label{fig:PQRSTYZ}
    \end{figure}

    \subsubsection{Why is it called PQRST Complex?}
    
    As the gait cycle exists in everyone's accelerometer signal and swing phase is part of the gait cycle, the fluctuation between point $P$ and $T$ exists in everyone's gait cycle. Furthermore, in our experiment, we find that the variation of the accelerometer signal between point $P$ and $T$ varies from person to person, but the overall structure remains the same. The overall structure means that everyone has the "flat-fluctuating-flat" shape for the whole gait cycle, and the "up-down-up-down" variation pattern persists in each fluctuating part. This structure reminds us of the "P wave, QRS complex, T wave" structure in ECG. Therefore, we call this structure we captured in the accelerometer signal as "PQRST Complex". As this structure is ubiquitous in everyone's gait cycle and varies from person to person, we conjecture that this structure is unique for different people and can be used for identifying people. Our results, shown in Section \ref{sec:performance}, prove that our speculation is correct. 
    
    The PQRST Complex contains three valleys and two peaks. As shown in the middle part of Figure \ref{fig:accX2PQRST}, they are marked with red points. From left to right, we mark $P$ (valley), $Q$ (peak), $R$ (valley), $S$ (peak), $T$ (valley) consequently. We also check if this structure exists in "Acc\_Y" and "Acc\_Z". Figure \ref{fig:PQRSTYZ} displays the plot for "Acc\_Y" and "Acc\_Z" from subject 3 on the top of each sub-figure. The zoomed repetitive patterns are shown below the accelerometer plot. We mark the PQRST Complex in each of them with red points. It shows that this structure can be captured in all three axes from accelerometer signals. 

    \subsubsection{How to detect PQRST Complex?}\label{sec:howto}
    
    To capture the PQRST Complex, we first find the starting point $P$. According to our experiment, the point "P" is the minimum point of "Acc\_Z" in a gait cycle. In general, the length of one gait cycle varies between 800 ms - 1400 ms. Thus, the minimum point in a sliding window of 800 ms - 1400 ms is $P$. For "Acc\_X" and "Acc\_Y", the point $P$ appears a little before the point $P$ appears in "Acc\_Z". So, we find the local minima within 10 ms before the time when point $P$ of "Acc\_Z" appears to get the point $P$ in "Acc\_X" and "Acc\_Y". After finding the point $P$, we can find local maxima and local minima in turns twice, to find the following points, which are $Q$, $R$, $S$, and $T$. 
    
    \subsection{Extracting and analyzing features in the PQRST Complex structure - Amplitude features convey more information}\label{sec:featureanalysis}
    
    \begin{figure*}[!ht]
    \begin{minipage}{0.33\linewidth}
    \centering
    \subfloat[Amplitude features on Acc\_X.]{\label{fig:AmpHistX}\includegraphics[width=\linewidth]{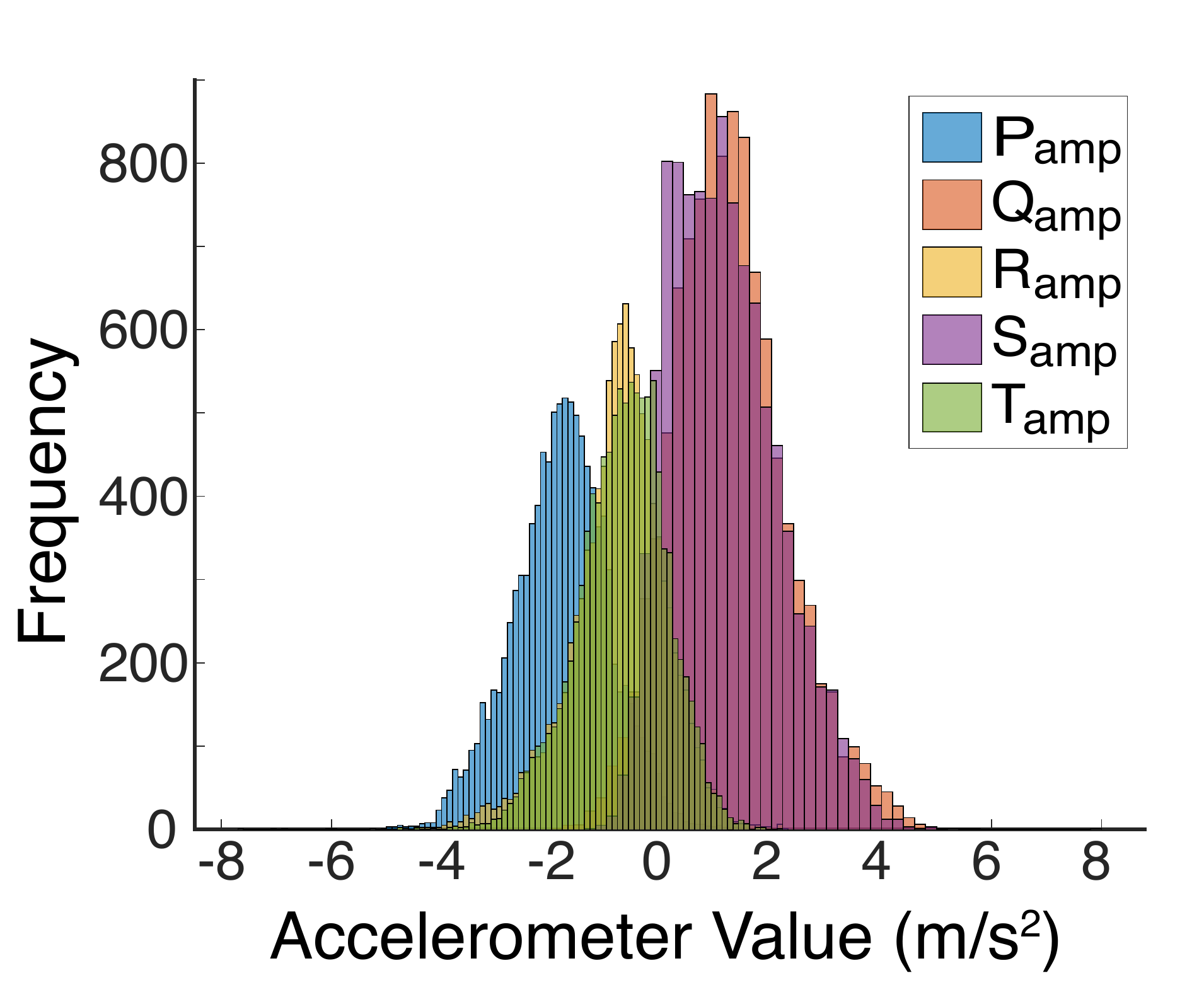}}
    \end{minipage}%
    \begin{minipage}{0.33\linewidth}
    \centering
    \subfloat[Amplitude features on Acc\_Y.]{%
          \includegraphics[width=\linewidth]{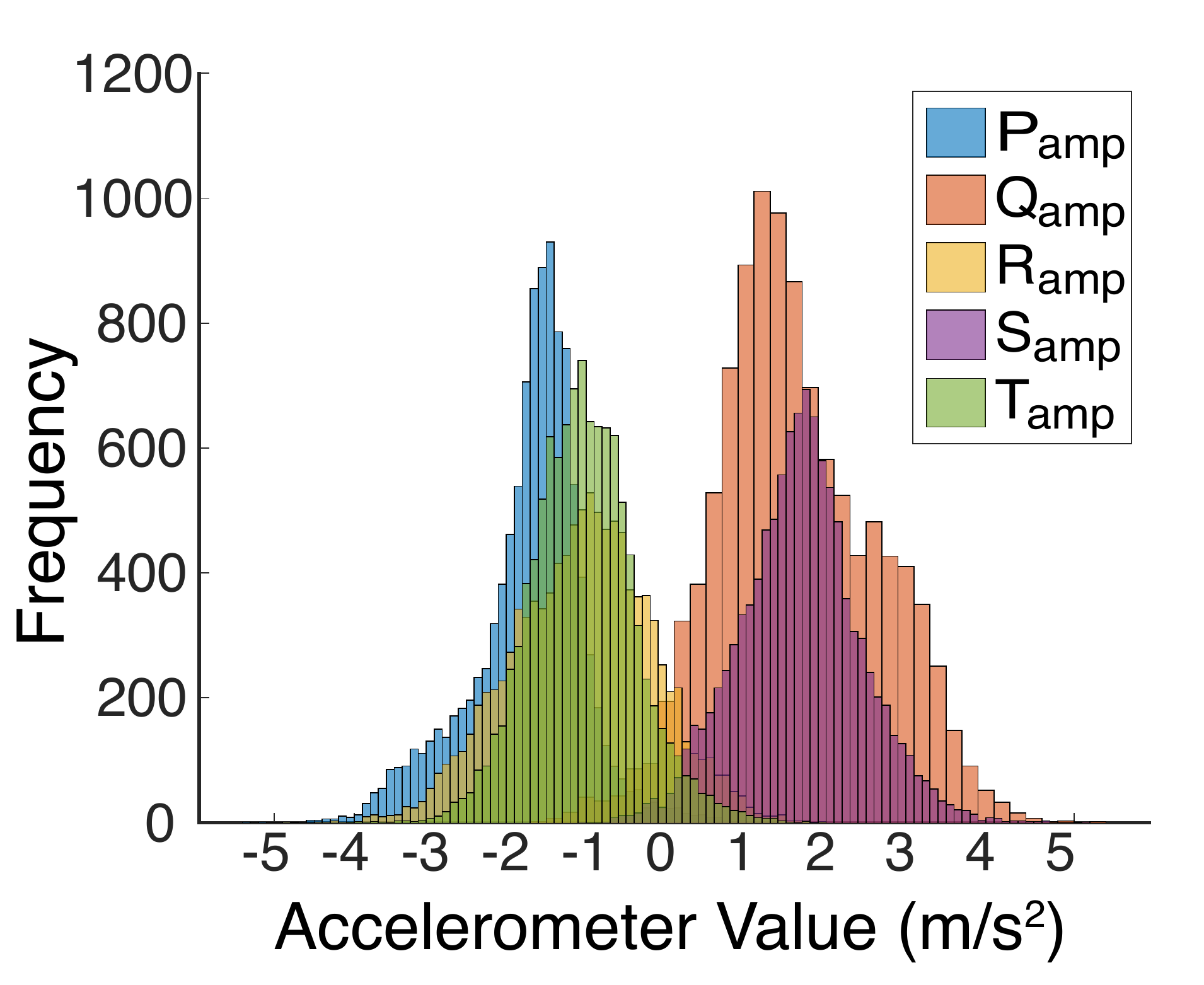}%
          \label{fig:AmpHistY}
        }
    \end{minipage}%
    \begin{minipage}{0.33\linewidth}
    \centering
    \subfloat[Amplitude features on Acc\_Z.]{%
          \includegraphics[width=\linewidth]{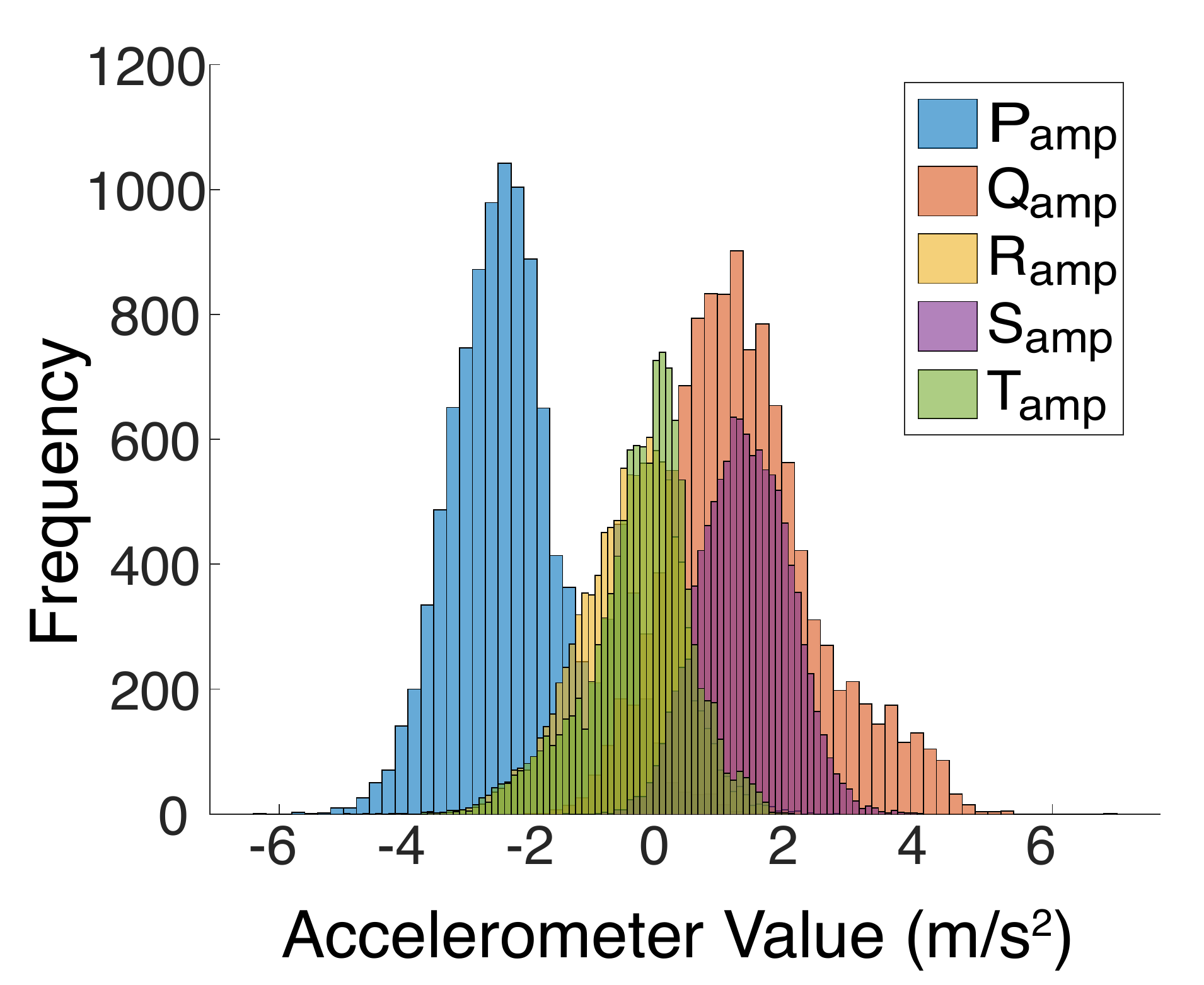}%
          \label{fig:AmpHistZ}
        }
    \end{minipage}%
    \par\medskip
    \begin{minipage}{0.33\linewidth}
    \centering
    \subfloat[Interval features on Acc\_X.]{%
          \includegraphics[width=\linewidth]{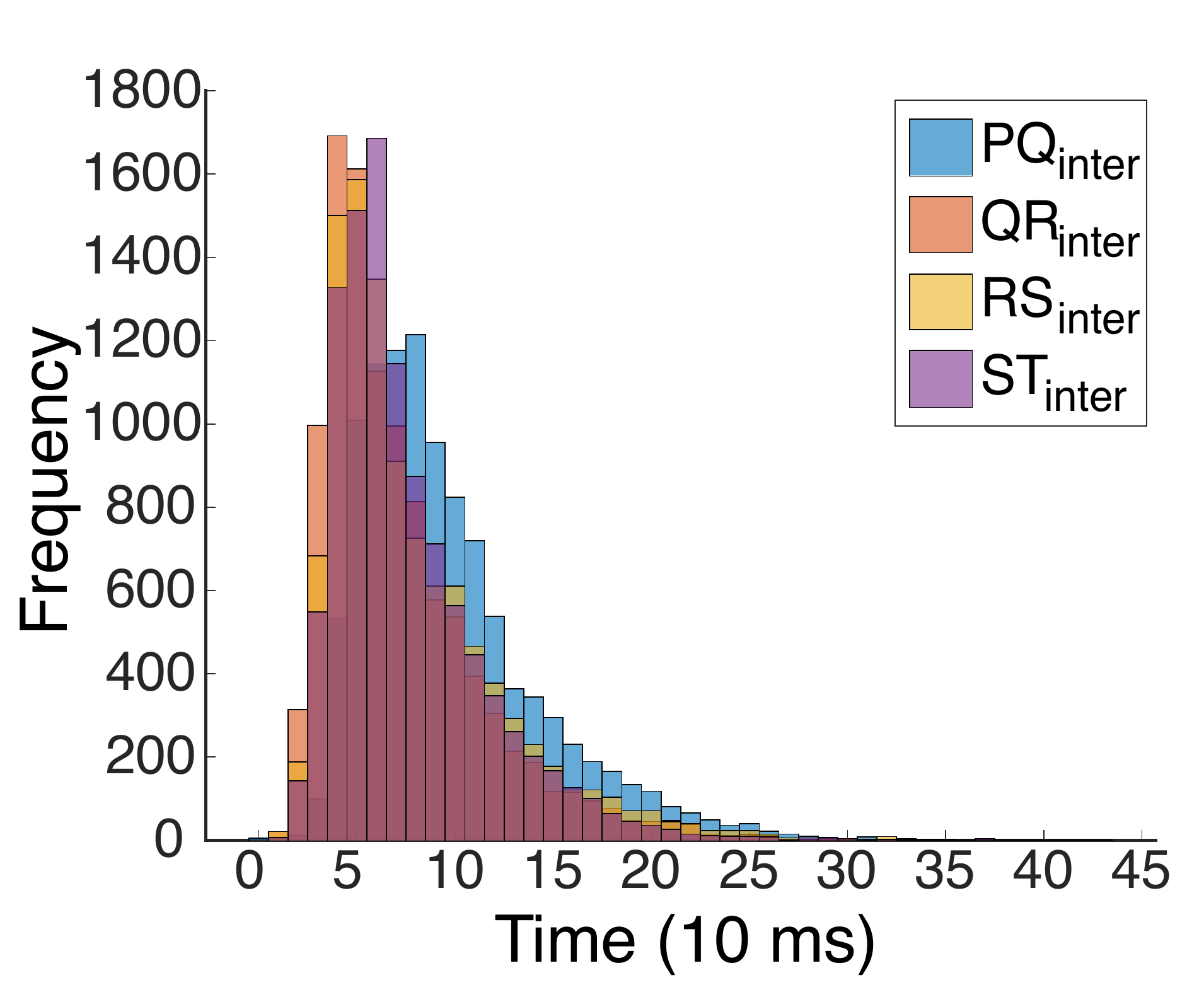}%
          \label{fig:IntHistX}
        }
    \end{minipage}%
    \begin{minipage}{0.33\linewidth}
    \centering
    \subfloat[Interval features on .]{%
          \includegraphics[width=\linewidth]{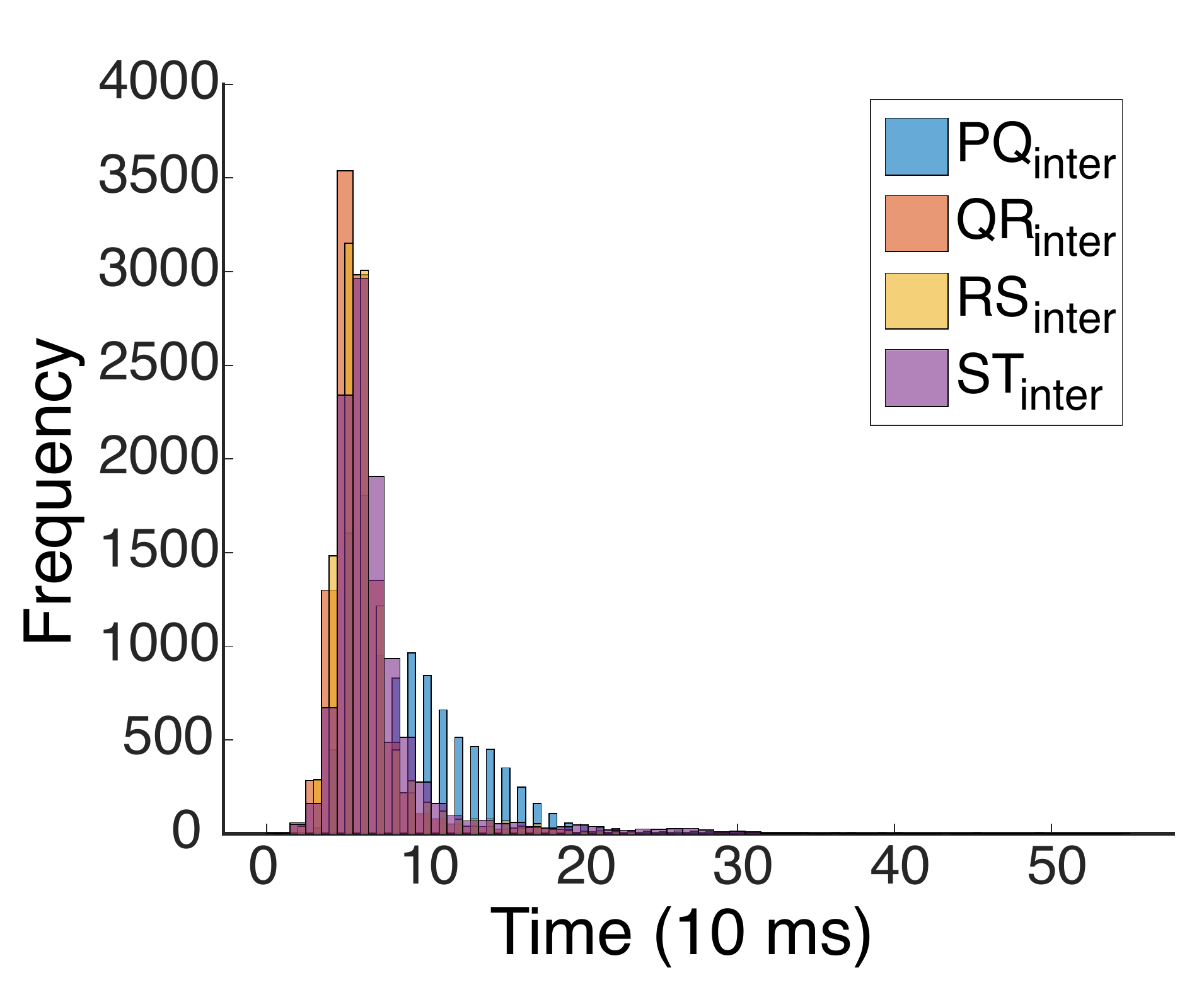}%
          \label{fig:IntHistY}
        }
    \end{minipage}%
    \begin{minipage}{0.33\linewidth}
    \centering
    \subfloat[Interval features on Acc\_Z.]{%
          \includegraphics[width=\linewidth]{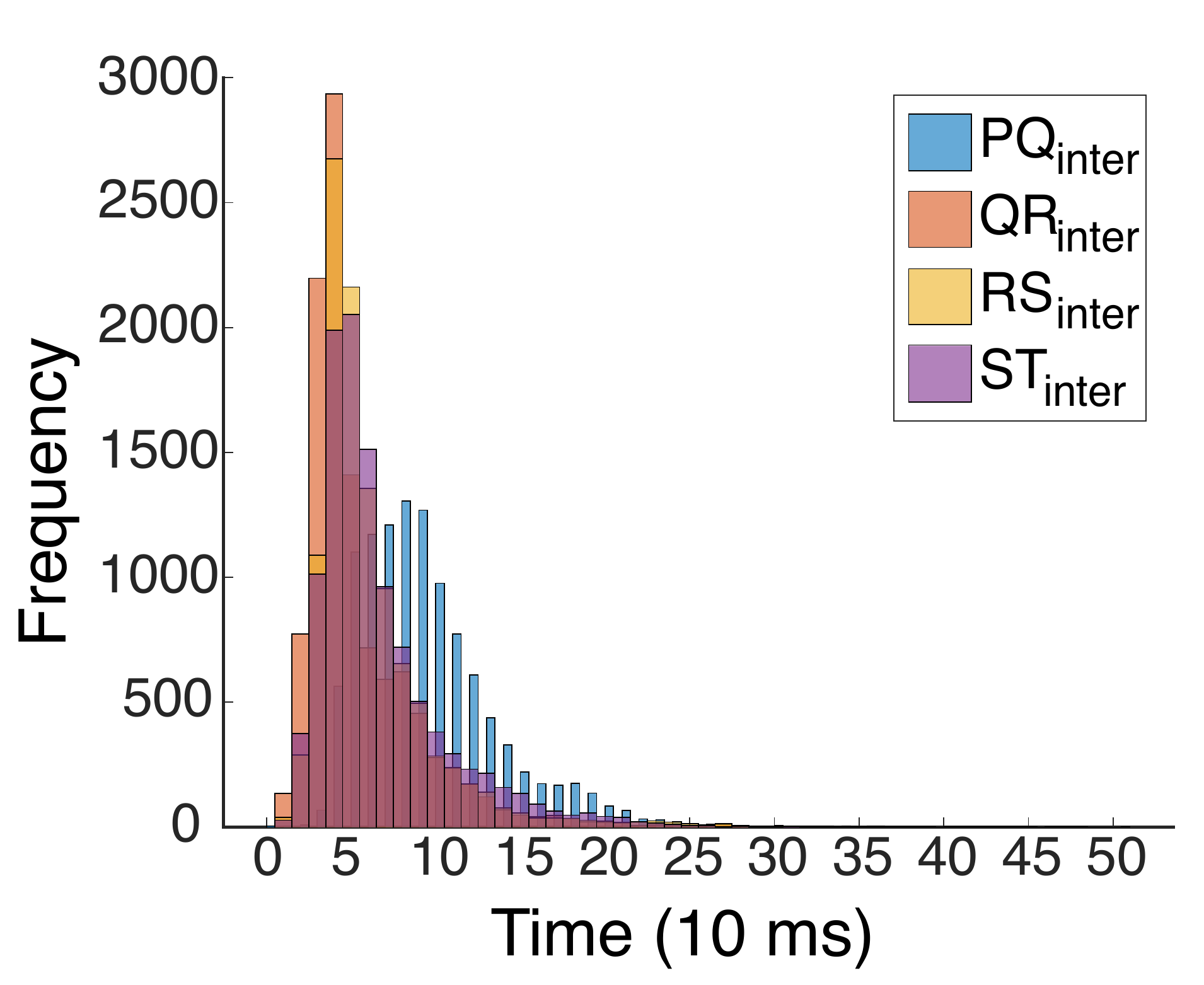}%
          \label{fig:IntHistZ}
        }
    \end{minipage}%
    \par\medskip
    \caption{Histogram plots for features from the PQRST complex. Figures in the upper row are amplitude features, and figures in the lower row are interval features. Amplitude features have wide spread of distributions. Interval features have narrow spread of distributions and are right skewed. Also, they share a large part of intersection. }
    \label{fig:Hist}
    \end{figure*}
    
    To describe this structure comprehensively, we define several features within the PQRST Complex in terms of amplitudes of acceleration values and time separations between foot events during the swing phase. These features are similar to the features used for ECG analysis. The amplitude of $P$ is expressed as $P_{amp}$, the amplitude of $Q$ is $Q_{amp}$, the amplitude of $R$ is $R_{amp}$, the amplitude of $S$ is $S_{amp}$ and the amplitude of $T$ is $T_{amp}$. These five amplitude features are acceleration values and unit is $m/s^2$. The interval between $P$ and $Q$ is $PQ_{inter}$, the interval between $Q$ and $R$ is $QR_{inter}$, the interval between $R$ and $S$ is $RS_{inter}$, the interval between $S$ and $T$ is $ST_{inter}$. The unit of these four interval features is millisecond. These nine features are marked in the middle figure of Figure \ref{fig:accX2PQRST}. 

    We divide these nine features into two feature sets, the amplitude feature set, and the interval feature set. These three feature sets are shown as follows, and we will use these names in the rest of the paper: 
    
    \begin{itemize}
        \item Feature set 1 (only includes five amplitude features): $P_{amp}$, $Q_{amp}$, $R_{amp}$, $S_{amp}$, $T_{amp}$.
        \item Feature set 2 (only contains four interval features): $PQ_{inter}$, $QR_{inter}$, $RS_{inter}$, $ST_{inter}$.
        \item Feature set 3 (contains all nine features): features set 1 + feature set 2.
    \end{itemize}

    Figure \ref{fig:Hist} shows the histograms of feature set 1 and feature set 2 on each axis from all subjects respectively. It is obvious that for interval features (feature set 2) on all three axes (Figure \ref{fig:IntHistX}, \ref{fig:IntHistY} and \ref{fig:IntHistZ}), the distributions are similar and largely overlapping. The distributions are not symmetric, but right skewed. Also, the spreads of the distributions are narrow, and they have a large part of intersection of each other. However, for amplitude features (feature set 1) (Figure \ref{fig:AmpHistX}, \ref{fig:AmpHistY} and \ref{fig:AmpHistZ}), features are distributed differently, centers are distinct and they do not share too many similar values. Also, the spreads of the distributions are wide, and the range for each feature is wider than interval features. It suggests that amplitude features may provide more classification information than interval features. Furthermore, features are more differently distributed on y-axis than those on the other two axes. In our experiments, we evaluate the performance of classification on these three feature sets and make a comparison among them. More details are illustrated in Section \ref{sec:exp}.

    \subsection{Importance and application of the PQRST Complex in gait analysis}
    
    The PQRST Complex structure is the first formally defined standard structure in gait cycles from accelerometer signals. It corresponds to the swing phase in gait cycles and captures the sequence of events during this phase. It matches the foot movement in the gait cycle, including the raising of the foot, flight of foot, and lowering of the foot. And this sequence is universal to any individual. Like the "QRS Complex" structure in ECG signals, the PQRST Complex provides a standardized structure for gait cycle representation. This structure persists in everyone's gait cycle, but its shape may vary from person to person. It leads us to a new direction of gait analysis. 
    
    Instead of investigating the whole gait cycle, we can focus on only this PQRST Complex since it contains most of the information of the whole gait cycle with fewer points and higher efficiency. Prior studies \cite{Derawi2010Improved, gafurov2007gaitauthen, gafurov2007spoof, gafurov2006robustness, Rong2007Identification} consider the gait cycle as a whole for recognition to avoid losing structural information. They detect the whole gait cycle first and create templates from several gait cycles. This usually requires a large amount of gait data for training and many gait cycles to get reliable templates. In contrast, we create one feature vector from each PQRST Complex, and get sufficient training samples with a short amount of gait data. The above studies use distance-based or similarity-based methods to compare input samples with stored templates in the authentication process. It takes a lot of time and space to compute the distance between about 100 pairs of points \cite{derawi2012smartphones, gafurov2007gaitauthen}, which is not efficient when applied to mobile devices. However, the feature vector we extract from our PQRST Complex has only nine dimensions. We can use simple classifiers to train the dataset and get a relatively high classification accuracy. And it only takes a small amount of time and space for training and testing, which is suitable for mobile phones. 
    
    Aggregate feature-based methods reduce the number of features used for authentication by extracting the maximum, minimum, median, etc., of the accelerometer readings. However, they miss critical structure information of the gait cycle. For instance, a gait cycle in which, the maximum appears in the first half should be different from the one in which, the maximum appears in the second half. But they are considered the same when only using maximum as features. Other aggregate features have similar issues. Thus, aggregate features are more likely to be imitated and may not appropriately represent the intricacies of the walking patterns. In contrast, our PQRST Complex focuses on the swing phase of the gait cycle, which retains as much important information as possible. At the same time, it keeps the structure information by extracting interval and amplitude features. Thus, our PQRST Complex eliminates the shortcomings of template-based methods and aggregate feature-based methods while retaining the information representing walking patterns. 

    Standardization of this structure paves the way for the comparison of multiple studies and future research. For example, it matches the variations in the accelerometer signal with actual walking events. The variations in the PQRST Complex may be attributed to the muscle-skeletal structure, the pace of walking, and other factors such as the carrying of objects and wearing coats. We can use it to estimate step length and leg length, estimate velocity, and do gait cycle detection, activity detection, and carry-on items detection.

\section{Experiments}\label{sec:exp}

    \subsection{Data description and data pre-processing} \label{sec:datadescription}

    The dataset we use for this paper is a public dataset from IEEE DataPort \cite{rpaz-0h66-19}. This dataset includes several types of biometric data from 117 subjects. It contains accelerometer and gyroscope for gait, typing  for keystroke dynamics, and swiping on mobile phones. In this paper, we only use accelerometer data collected during walking tasks to test our authentication system. 
    
    Our paper focuses on the walking pattern when people walk on flat ground, so we only use data from the following three tasks: Task 1, subjects leave the lab and walk in the corridor. Task 3, subjects walk in another corridor. Task 5, subjects walk in the first corridor again and return to the lab. Task 1 and 5 last around 20 seconds, and Task 3 lasts around 60 seconds. Subjects are asked to perform the whole walking process twice on two separate days. Data collected on the first day are called Session 1, and those collected on the second day are called Session 2. During the whole walking process, subjects are asked to put a mobile phone in the pocket of their pants, and the built-in accelerometer keeps track of the acceleration force in $m/s^2$ on x, y, and z axis, respectively, excluding the force of gravity. The frequency of the data is 100 Hz, so 100 data points represent one second. 
    
    For data pre-processing, we first check all subjects' data and remove those with incomplete data. Then according to the checkpoints, we only retain the data for Tasks 1, 3, and 5 from the remaining dataset and remove those with less than 400 samples (equivalent to 4 seconds) in each task. Finally, 89 subjects are left. After that, we normalize the data with center 0 and standard deviation 1 using the z-score method for each axis. To remove noise, we smooth the data by using a sliding window to compute the average over the elements within each window. The length of the window is set to be 4 points.   
    
    To construct the training and testing dataset, we use data from Session 1 for training, and data from Session 2 are used for testing. For each subject, its training and testing sets are considered genuine samples and labeled as 0, while samples from other subjects are considered imposter samples and labeled as 1.

    \begin{table*}[!t]
    \centering
    \begin{tabular}{c|c|c|c|c|c|c}
    \hline
       & \multicolumn{2}{c|}{Feature Set 1} & \multicolumn{2}{c|}{Feature Set 2}                        & \multicolumn{2}{c}{Feature Set 3}                       \\ \cline{2-7} 
        & LDA      & SVM   & LDA       & SVM   & LDA      &  SVM   \\ \hline \hline
        Acc\_X & 81.26 (18.72) & 81.37 (18.61) & 74.97 (25.15) & 75.90 (24.25) & 83.48 (16.40) & 83.12 (16.79) \\ \hline
        Acc\_Y & 87.47 (12.66) & 87.70 (12.42)) & 85.00 (15.11)  & 85.33 (14.77) & 90.12 (9.97) & 90.51 (9.58) \\ \hline
        Acc\_Z & 86.74 (13.45) & 86.68 (13.45) & 80.05 (20.15) & 81.41 (18.76) & 89.43 (10.65) & 89.06 (11.02) \\ \hline
    \end{tabular}
    \caption{The average CCR and EER (numbers in the bracket) for each axis and each feature set of two classifier. All numbers are in \%. Feature Set 1 includes all five amplitude features, Feature Set 2 includes all four interval features, and Feature Set 3 includes all nine features. }
    \label{tb: results}
    \end{table*}

    \begin{table*}[h]
    \centering
    \begin{tabular}{c|c|c|c|c}
    \hline
                              & Classification methods & No. subjects & No. features        & Performance \\ \hline
    Gafurov \etal \cite{gafurov2007gaitauthen}     & Absolute distance      & 50                 & 100  & 86.3\% (CCR)  \\ \hline
    Derawi \etal \cite{Derawi2010Improved}        & DTW \& CRM             & 60                 & 100  & 5.7\% (EER)   \\ \hline
    Rong \etal \cite{Rong2007Identification}    & DTW                    & 35                 & Not mentioned             & 6.7\% (EER)   \\ \hline
    Thang \etal \cite{thang2012gait}    & DTW   & 11    & around 40     & 79.1\% (CCR) \\ \hline
    Nickel \etal \cite{nickel2012accelerometer}   & HMMs                   & 48                 & 52                        & 15.46\% (EER) \\ \hline \hline
    Our work                  & LDA,SVM                & 89                 & 9                         & 90.51\% (CCR) / 9.58\% (EER) \\ \hline
    \end{tabular}
    \caption{Performance comparison between our method and some prior studies. (DTW: Dynamic Time Warping, CRM: Cyclic rotation metric, HMMs: Hidden Markov Models)}
    \label{tb:compareResults}
    \end{table*}
    \subsection{Accelerometer-based gait authentication system using PQRST Complex}
    
     The authentication system first detects PQRST Complex from the training set using the method illustrated in Section \ref{sec:howto}. Then it extracts nine features and compose feature vector for each cycle.  It trains a Linear Discriminant Analysis (LDA) classifier and a Support Vector Machine (SVM) classifier separately for each pair of subjects with three feature sets, as we mentioned in Section \ref{sec:featureanalysis}. We use the Correct Classification Rate (CCR) and Equal Error Rate (EER) to evaluate the performance of our authentication system.
    
    \subsection{Performance evaluation and interesting observations}\label{sec:performance}
    
    As shown in Table \ref{tb: results}, our authentication system achieves 90.51\% CCR when using nine features with the SVM classifier, and it gets a similar performance when using the LDA classifier. It  suggests that we should include all nine features in our authentication system to make it more accurate. These results also show that even though the PQRST Complex structure only captures the swing phase of the gait cycle, it retains enough gait information for authentication and recognition. 
    
    Table \ref{tb:compareResults} shows the performance comparison between our work and some prior studies who used gait cycles. It is worth noting that we use the least amount of features to get a relatively high CCR and low EER among these studies. At the same time, the dataset we use to evaluate our authentication system is the largest, and those two studies that achieved lower EER than ours used a much smaller dataset. These results prove that our PQRST Complex structure can capture each individual's walking patterns and be used for gait authentication. With fewer features and less computation time, our PQRST Complex can still perform reasonably well in gait authentication systems compared to prior studies. It provides a new way of accelerometer-based gait analysis except for using templates of gait cycles, aggregate features, and raw data learned by neural networks. 
    
    Additionally, there are two interesting observations that we  find from the results. 
    
    \textbf{Observation 1: Amplitude features perform better than interval features in classification -- }
    The Feature set 1 has better performance than the Feature set 2 (Table \ref{tb: results} column "Feature set 1" and column "Feature set 2"). This result is the same as what we conjectured in Section \ref{sec:featureanalysis}, that amplitude features are more reliable than interval features in classifying subjects due to their ranges and distributions. It implies that only using amplitude features (Feature set 1) can still achieve a relatively high CCR compared to using nine features and further reduce the computational complexity. 
    
    \textbf{Observation 2: "Acc\_Y" and "Acc\_Z" have better performance than "Acc\_X" -- } 
    It is worth noting that the performance of "Acc\_Y" and "Acc\_Z" is better than "Acc\_X". When we put the cell phone in the pants pocket, the x-axis of the accelerometer in the cell phone is along the forward direction, the y-axis of the accelerometer is vertical to the ground, and the z-axis of the accelerometer is vertical to the body. These results show that people's walking pattern is not obvious in the forward direction, while these patterns are easier to find along y and z directions.

\section{Conclusion}\label{sec:conclusion}

    In this paper, we capture a new structure PQRST Complex in gait cycles from the accelerometer signal. This structure formally matches the sequence of gait events, including raising of the foot, flight of the foot, and lowering of the foot during the swing phase in a walking cycle. It can be captured from everyone's gait cycle and varies from person to person. We posit that our PQRST Complex can become a standardized structure to represent gait cycle and opens a new area for gait analysis. Analogous to the features in ECG structure, we present nine features within PQRST Complex and use them  for recognition and authentication. We construct an accelerometer-based gait authentication system using this structure and evaluate its performance using CCR and EER. Our authentication system outperforms many state-of-the-art gait cycle-based authentication systems with an average CCR of 90.51\%, while average EER is only 9.58\%. The experimental results show that the PQRST Complex structure and features contain sufficient gait information for authentication. Compared to other studies, our authentication system not only performs well on a larger dataset, but it requires fewer features, lesser computation time and space, and shorter training data. 

{\small
\bibliographystyle{ieee}
\bibliography{reference}

\begin{thebibliography}{10}\itemsep=-1pt

\bibitem{ailisto2005identifying}
H.~Ailisto, M.~Lindholm, J.~Mäntyjärvi, E.~Vildjiounaite, and S.-M. Mäkelä.
\newblock Identifying people from gait pattern with accelerometers.
\newblock {\em Proc SPIE}, 5779, 03 2005.

\bibitem{alobaidi2022real}
H.~Alobaidi, N.~Clarke, F.~Li, and A.~Alruban.
\newblock Real-world smartphone-based gait recognition.
\newblock {\em Computers \& Security}, 113:102557, 2022.

\bibitem{annadhorai2008human}
A.~Annadhorai, E.~Guenterberg, J.~Barnes, K.~Haraga, and R.~Jafari.
\newblock Human identification by gait analysis.
\newblock In {\em Proceedings of the 2nd International Workshop on Systems and
  Networking Support for Health Care and Assisted Living Environments}, pages
  1--3, 2008.

\bibitem{anwary2021Insolebased}
A.~R. Anwary, D.~Arifoglu, M.~Jones, M.~Vassallo, and H.~Bouchachia.
\newblock Insole-based real-time gait analysis: Feature extraction and
  classification.
\newblock In {\em 2021 IEEE International Symposium on Inertial Sensors and
  Systems (INERTIAL)}, pages 1--4, 2021.

\bibitem{anwary2018automatic}
A.~R. Anwary, H.~Yu, and M.~Vassallo.
\newblock An automatic gait feature extraction method for identifying gait
  asymmetry using wearable sensors.
\newblock {\em Sensors}, 18(2):676, 2018.

\bibitem{bejarano2014novel}
N.~C. Bejarano, E.~Ambrosini, A.~Pedrocchi, G.~Ferrigno, M.~Monticone, and
  S.~Ferrante.
\newblock A novel adaptive, real-time algorithm to detect gait events from
  wearable sensors.
\newblock {\em IEEE transactions on neural systems and rehabilitation
  engineering}, 23(3):413--422, 2014.

\bibitem{derawi2012smartphones}
M.~O. Derawi.
\newblock Smartphones and biometrics: Gait and activity recognition.
\newblock 2012.

\bibitem{Derawi2010Improved}
M.~O. Derawi, P.~Bours, and K.~Holien.
\newblock Improved cycle detection for accelerometer based gait authentication.
\newblock In {\em 2010 Sixth International Conference on Intelligent
  Information Hiding and Multimedia Signal Processing}, pages 312--317, 2010.

\bibitem{gafurov2007survey}
D.~Gafurov.
\newblock A survey of biometric gait recognition: Approaches, security and
  challenges.
\newblock In {\em Annual Norwegian computer science conference}, pages 19--21.
  Annual Norwegian Computer Science Conference Norway, 2007.

\bibitem{gafurov2007gaitauthen}
D.~Gafurov, E.~Snekkenes, and P.~Bours.
\newblock Gait authentication and identification using wearable accelerometer
  sensor.
\newblock In {\em 2007 IEEE Workshop on Automatic Identification Advanced
  Technologies}, pages 220--225, 2007.

\bibitem{gafurov2007spoof}
D.~Gafurov, E.~Snekkenes, and P.~Bours.
\newblock Spoof attacks on gait authentication system.
\newblock {\em IEEE Transactions on Information Forensics and Security},
  2(3):491--502, 2007.

\bibitem{gafurov2006robustness}
D.~Gafurov, E.~Snekkenes, and T.~E. Buvarp.
\newblock Robustness of biometric gait authentication against impersonation
  attack.
\newblock In {\em OTM Confederated International Conferences" On the Move to
  Meaningful Internet Systems"}, pages 479--488. Springer, 2006.

\bibitem{hou2014systemforhumangait}
J.~{Hou}, R.~{Ji}, C.~{Qin}, Y.~{Yang}, C.~{Wang}, and Z.~{Wang}.
\newblock A system for human gait analysis based on body sensor network.
\newblock In {\em 2014 International Conference on Wireless Communication and
  Sensor Network}, pages 343--347, 2014.

\bibitem{lee2002gait}
L.~Lee and W.~E.~L. Grimson.
\newblock Gait analysis for recognition and classification.
\newblock In {\em Proceedings of Fifth IEEE International Conference on
  Automatic Face Gesture Recognition}, pages 155--162. IEEE, 2002.

\bibitem{mantyjarvi2005identifyingusers}
J.~{Mantyjarvi}, M.~{Lindholm}, E.~{Vildjiounaite}, S.~. {Makela}, and H.~A.
  {Ailisto}.
\newblock Identifying users of portable devices from gait pattern with
  accelerometers.
\newblock In {\em Proceedings. (ICASSP '05). IEEE International Conference on
  Acoustics, Speech, and Signal Processing, 2005.}, volume~2, pages
  ii/973--ii/976 Vol. 2, 2005.

\bibitem{Maqbool2017realtime}
H.~F. Maqbool, M.~A.~B. Husman, M.~I. Awad, A.~Abouhossein, N.~Iqbal, and A.~A.
  Dehghani-Sanij.
\newblock A real-time gait event detection for lower limb prosthesis control
  and evaluation.
\newblock {\em IEEE Transactions on Neural Systems and Rehabilitation
  Engineering}, 25(9):1500--1509, 2017.

\bibitem{mufandaidza2018continuous}
M.~Mufandaidza, T.~Ramotsoela, and G.~Hancke.
\newblock Continuous user authentication in smartphones using gait analysis.
\newblock In {\em IECON 2018 - 44th Annual Conference of the IEEE Industrial
  Electronics Society}, pages 4656--4661, 2018.

\bibitem{neumann2010kinesiology}
D.~Neumann.
\newblock {\em Kinesiology of the Musculoskeletal System: Foundations for
  Rehabilitation}.
\newblock Volve learning system. Mosby/Elsevier, 2010.

\bibitem{nickel2012accelerometer}
C.~Nickel.
\newblock Accelerometer-based biometric gait recognition for authentication on
  smartphones.
\newblock 2012.

\bibitem{niyogi1994analyzing}
S.~A. Niyogi and E.~H. Adelson.
\newblock Analyzing gait with spatiotemporal surfaces.
\newblock In {\em Proceedings of 1994 IEEE Workshop on Motion of Non-rigid and
  Articulated Objects}, pages 64--69. IEEE, 1994.

\bibitem{rpaz-0h66-19}
A.~K. B. L. W. S. S. I. P. S. R. W. R. D. J. B. Z. J. V.~V. Phoha.
\newblock Su-ais bb-mas (syracuse university and assured information security -
  behavioral biometrics multi-device and multi-activity data from same users)
  dataset, 2019.

\bibitem{Rong2007Identification}
L.~Rong, D.~Zhiguo, and Z.~Jianzhong.
\newblock Identification of individual walking patterns using gait
  acceleration.
\newblock pages 543 -- 546, 08 2007.

\bibitem{tamviruzzaman2009epetwhen}
M.~Tamviruzzaman, S.~I. Ahamed, C.~S. Hasan, and C.~O'brien.
\newblock Epet: When cellular phone learns to recognize its owner.
\newblock In {\em Proceedings of the 2nd ACM Workshop on Assurable and Usable
  Security Configuration}, SafeConfig '09, page 13–18, New York, NY, USA,
  2009. Association for Computing Machinery.

\bibitem{thang2012gait}
H.~M. Thang, V.~Q. Viet, N.~D. Thuc, and D.~Choi.
\newblock Gait identification using accelerometer on mobile phone.
\newblock In {\em 2012 International Conference on Control, Automation and
  Information Sciences (ICCAIS)}, pages 344--348. IEEE, 2012.

\bibitem{wan2018survey}
C.~Wan, L.~Wang, and V.~V. Phoha.
\newblock A survey on gait recognition.
\newblock {\em ACM Computing Surveys (CSUR)}, 51(5):1--35, 2018.

\end{thebibliography}
}

\end{document}